\documentclass[aps,prl,reprint,letterpaper,superscriptaddress,showpacs]{revtex4-1}
\usepackage{CJK}
\usepackage{keyval}%
\usepackage{graphicx}
\usepackage{dcolumn}
\usepackage{bm}
\usepackage{color}

\setkeys{Gin}{width=0.95\columnwidth,clip=true}
\newcommand{\ignore}[1]{}

\newcommand{\material}{K$_{0.8}$Fe$_{1.6}$Se$_2$}

\begin{document}
\begin{CJK*}{UTF8}{}
% You may use Title,Subject,Author,Manager,Company,Operator,
% Category,Comment,Hlinkbase document properties here
\title{Novel Insulating Magnetism in Vacancy-Ordered K$_{0.8}$Fe$_{1.6}$Se$_2$}
%\author{Wei-Guo Yin
\author{Wei-Guo Yin (\CJKfamily{gbsn}尹卫国)
}
\email[Corresponding author; ]{wyin@bnl.gov}%
%\email{wyin@bnl.gov}%
\affiliation{Condensed Matter Physics and Materials Science Department,
Brookhaven National Laboratory, Upton, New York 11973, USA}
%\author{Chia-Hui Lin
\author{Chia-Hui Lin (\CJKfamily{bsmi}林佳輝)
}
%\author{Wei Ku
\author{Wei Ku (\CJKfamily{bsmi}顧威)
}
\affiliation{Condensed Matter Physics and Materials Science Department,
Brookhaven National Laboratory, Upton, New York 11973, USA}
\affiliation{Department of Physics and Astronomy, Stony Brook University,
Stony Brook, New York 11794, USA}

\date{\today}

\begin{abstract}
We unveil the novel physical origin of the insulating block checkerboard antiferromagnetism in vacancy-ordered {\material}. Our first-principles electronic structure analysis reveals its incompatibility with a simple Fermi-surface nesting or Mott insulator scenario, and suggests the picture of coexisting itinerant and localized electronic states. Consistently, we demonstrate that it can be unified with the metallic collinear or bicollinear antiferromagnetism of the vacancy-free parent compounds LaOFeAs, BaFe$_2$As$_2$, or FeTe in the spin-fermion model. These results indicate that the blocking effects of Hund's rule coupling and the resulting electron correlation are crucial to the electronic and magnetic structures of iron-based superconductors.

\end{abstract}
\pacs{
%74. Superconductivity
%74.25.Jb,%Electronic structure
%74.25.Ha,%Magnetic properties
%74.20.Mn,%Nonconventional mechanisms
74.70.Xa,%Pnictides and chalcogenides
75.10.-b,%General theory and models of magnetic ordering
71.27.+a,%Strongly correlated electron systems; heavy fermions
%71.15.Mb,%Density functional theory, local density approximation, gradient and other corrections
%71.15.Nc%Total energy and cohesive energy calculations
75.25.Dk%Orbital, charge, and other orders, including coupling of these orders
}

\maketitle
\end{CJK*}

One of the puzzling phenomena in iron-based superconductors (FeSCs) is that despite apparent similarity in crystal and electronic structures, their parent compounds exhibit different metallic antiferromagnetic (AF) patterns: collinear $C$-type (Fig. 1a) and bicollinear $E$-type (Fig. 1b) in pnictides \cite{neu:1111:Cruz,neu:122:Huang} and chalcogenides \cite{neu:11:Bao}, respectively. Moreover, a transition between the $C$ and $E$ types in the same material can be induced by merely varying the anion height from the Fe plane \cite{dft:11:Moon}. Such magnetic softness implies the presence of strong spin fluctuations and electronic correlation, which are generally believed to be at the heart of the high-$T_c$ mechanism \cite{review:Mazin,_Mott:Si}. It is thus urgent to resolve this puzzle and to classify the essential nature of electronic correlation in FeSCs.

Recently, a new horizon to look into these problems emerges with the discovery of the $A_{1-y}$Fe$_{2-x}$Se$_2$ family of FeSCs \cite{syn:se122:Guo,syn:se122:Fang}, where the considerable amount of Fe vacancies induce substantial changes in electronic and magnetic structures \cite{[{For review, see }]review:se122:Mazin}. In particular, the parent compound {\material} exhibits an unusual insulating $2\times 2$ block checkerboard AF order (Fig.~1f, referred to \textit{X}-type from now on) \cite{neu:K0.8Fe1.6Se2:Bao,neu:se122:Wang,x-ray:se122:Ricci}. The 20\% Fe vacancies in it form a $\sqrt{5}\times\sqrt{5}$ order below $T_S$=578 K, on top of which the \textit{X}-type spin order develops below $T_N$=559 K with a large ordered Fe magnetic moment $3.3$ $\mu_B$. In a broad perspective, {\material} has brought in an ideal benchmark against magnetic theories for FeSCs: With fixed parameters good for vacancy-free chalcogenides, introducing the ordered vacancies should transit the metallic $E$ type to the insulating $X$ type. In particular, understanding the metal-insulation transition will yield insight into how electrons become correlated in FeSCs.

%\begin{figure*}[b]
\begin{figure}[b]
%\vspace{1cm}
\includegraphics[width=\columnwidth,clip=true,angle=0]{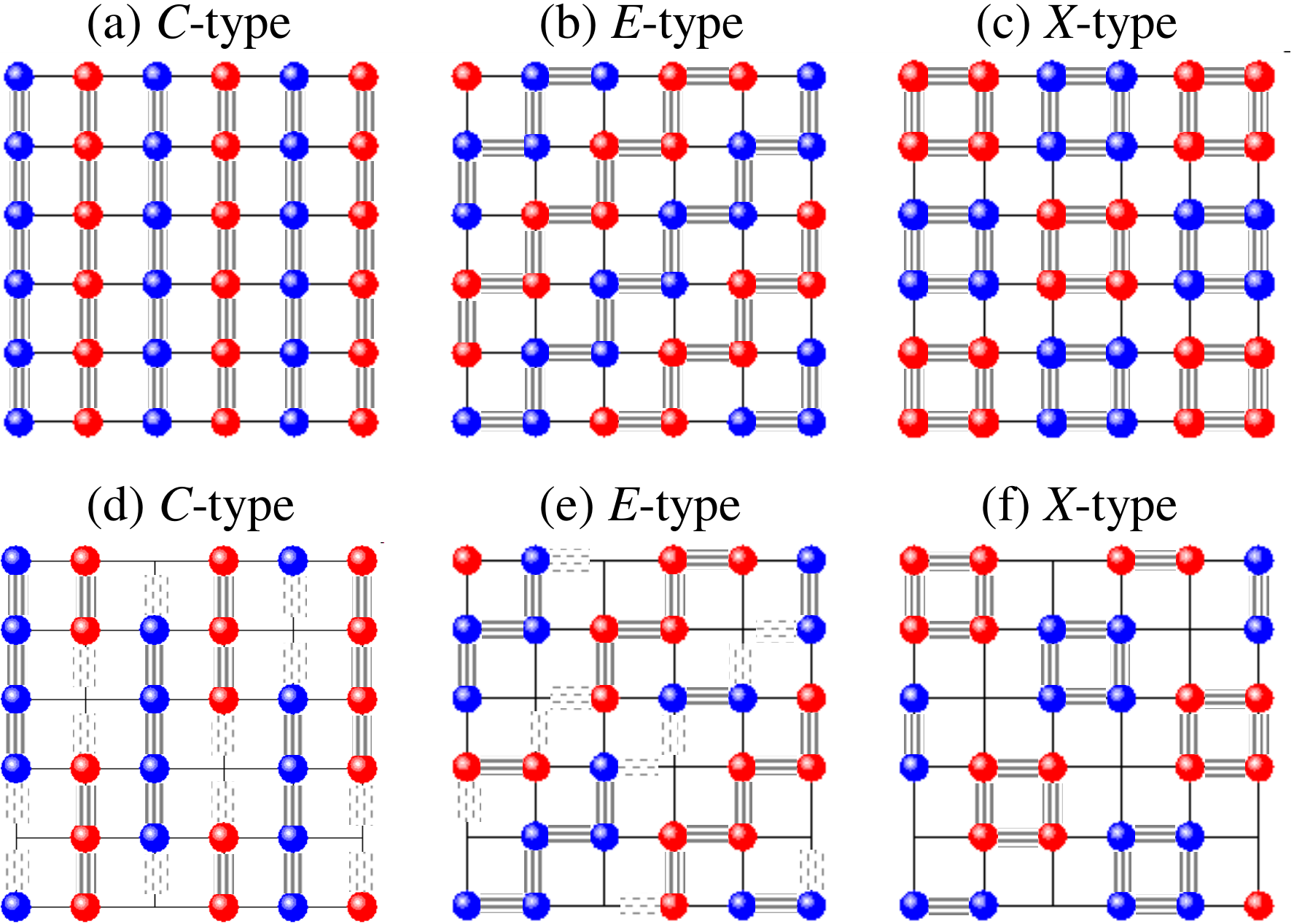}
\caption{\label{fig1} The in-plane patterns of the spin-up (blue
balls) and spin-down (red balls) iron atoms in (a) collinear \textit{C}-type, (b) bicollinear \textit{E}-type, and (c) block checkerboard \textit{X}-type AF states. Their counterparts in the presence of $\sqrt{5}\times\sqrt{5}$ Fe vacancy ordering are (d)-(f),  respectively. Patterned horizontal and vertical solid lines represent $d_{yz}$ and $d_{xz}$ bonds, respectively; dashed lines denote the Fe vacancy broken bonds.}
\end{figure}
%\end{figure*}

So far, the weakly interacting itinerant-electron model based on the experimentally observed Fermi surface topology can reproduce the $C$ type only \cite{review:Mazin,_weak:Graser,_weak:Knolle,note:Eremin}. The Heisenberg spin model, as fit to neutron scattering data on CaFe$_2$As$_2$ \cite{neu:122:Zhao}, FeTe \cite{neu:11:Lipscombe}, and {\material}  \cite{neu:se122:Wang}, shows that the $E$ and $X$ types share similar model parameters, but the $C$ and $E$ types are surprisingly well separated in the model parameter space, with the leading exchange interaction being AF and ferromagnetic (FM), respectively. To recover the $C-E$ proximity and their metallicity, a model with coexisting itinerant electrons and localized spins was proposed \cite{_DE:11:Yin}. Interestingly, all these spin orders have been reproduced in first-principles band calculations \cite{dft:1111:Yildirim,dft:11:Ma,dft:K0.8Fe1.6Se2:Yan,dft:KFe1.6Se2:Cao}; however, the microscopic origin of the magnetic softness and the metal-insulating transition remains to be elucidated. For example, the insulating nature of {\material} has been debated between Mott insulator \cite{_strong:se122:Yu} and magnetic semiconductor \cite{dft:K0.8Fe1.6Se2:Yan} scenarios. It is thus important and timely to carefully examine the electronic structure of vacancy-ordered {\material} towards a unified picture.

In this Letter, we present a combined first-principles and effective Hamiltonian analysis of the electronic structure of vacancy-ordered {\material}. Our first-principles results show (i) that its nonmagnetic Fermi surface does not display a nesting vector, and the bare spin susceptibility $\chi_0(\mathbf{q},\omega=0)$ is rather featureless and considerably weaker than the vacancy-free cases, rendering the Fermi surface instability an unlikely scenario, and (ii) that its magnetic electronic structure features two ``gaps'': a high-energy Mott gap and a low-energy Fe-Fe bonding gap, suggesting the picture of coexisting itinerant and localized electronic states \cite{osmt:Medici,_DE:1111:Kou,_DE:Lv,_DE:11:Yin,_DE:K0.8Fe1.6Se2:You}. Consistently, we demonstrate that the insulating antiferromagnetism of {\material} can be unified with the metallic antiferromagnetism of vacancy-free FeSCs in the spin-fermion model \cite{_DE:11:Yin}. These findings indicate that the blocking effects of Hund's rule coupling at low-energy scale and the resulting electronic correlation are crucial to the electronic and magnetic structures of FeSCs \cite{_strong:dmft:Yin,_strong:dmft:Yin_NM}.

%\begin{figure*}[b]
\begin{figure}[b]
%\vspace{1cm}
\includegraphics[width=\columnwidth,clip=true,angle=0]{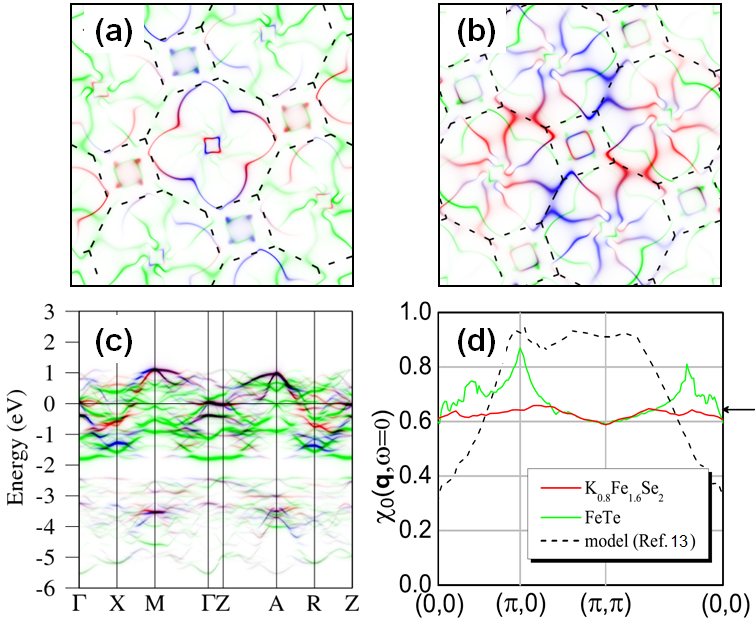}
\caption{\label{fig1:Fermi surface} LDA Fermi surface at $k_z$=0 (a) and $k_z$=$\pi$ (b) and band structure (c) in the one-Fe unit cell notation. High intensity means high spectral weight. Shadow bands appear as shifting the main bands by $(\pi,\pi,\pi)$ and $\mathbf{Q}=\pm(3\pi/5,\pi/5,\pi)$, $\pm(-\pi/5,3\pi/5,\pi)$. Color coded for $d_{xz}$ (red), $d_{yz}$ (blue), and the rest $3d$ orbitals (green). Dashed lines in (a),(b) are the folded Brillouin zone boundary in the in-plane 8-Fe unit cell notation. (d) Bare spin susceptibility in the one-Fe unit cell notation, with the value at $\mathbf{Q}$ indicated by the arrow, compared with those of FeTe and a previous model ~\cite{_weak:Graser}.
}
\end{figure}
%\end{figure*}

\textsl{First-Principles Analysis}~\cite{note:supplement}.---Since nonmagnetic {\material} is metallic \cite{dft:K0.8Fe1.6Se2:Yan}, a first question that needs to be clarified is whether the magnetic and metal-insulator transitions in {\material} is driven by Fermi surface instability. To this end, the electronic structure of {\material} at the experimental crystal structure \cite{se122:xray:Zavalij} was calculated within local density approximation (LDA) of density functional theory (DFT), implemented via full potential, all-electron, linearized augmented plane wave basis \cite{wien2k}. The resulting Fermi surface (more precisely the intensity of the one-particle propagator at Fermi level) unfolded \cite{,dft:Ku} into the one-Fe unit cell notation is presented in Figs.~2(a) and 2(b) \cite{[{For technical details, see }]dft:Lin}. No Fermi surface nesting is visible at $(3\pi/5, \pi/5)$ and $(-\pi/5, 3\pi/5)$, the characteristic wavevectors of the $X$-type spin order.

To be quantitative, we calculated the bare spin susceptibility $\chi_0(\mathbf{q},\omega=0)$ and unfolded it into the one-Fe unit cell notation \cite{note:supplement}. In comparison, we also calculated $\chi_0(\mathbf{q},0)$ for two vacancy-free cases having the Fermi surface nesting between hole pockets around $(0,0)$ and electron pockets around $(\pi,0)$ and $(0,\pi)$: a one-Fe-unit-cell model \cite{_weak:Graser} and FeTe in LDA. As shown in Fig.~2(d), the real part of $\chi_0(\mathbf{q},0)$ in {\material} is rather featureless, in sharp contrast with the vacancy-free cases. Besides, its maximum value is substantially smaller, making it much harder for the interaction-assisted instability \cite{_weak:Graser} to be effective. Hence, Fermi surface instability is unlikely a driving force for the magnetism in iron chalcogenides.

Next, we elucidate the nature of the band gap opening in {\material}. On the one hand, as shown in Fig.~2(c), removing one quarter of nearest neighbors in the Fe plane (Fig.~1f) reduces the bandwidth $W$ of nonmagnetic {\material} from $\sim4$ eV in vacancy-free FeSCs \cite{dft:Lin,dft:1111:Lee} to $\sim3$ eV; thus, the larger $U/W$ might favor the Mott metal-insulator transition, as illustrated in a two-orbital model without Hund's rule coupling \cite{_strong:se122:Yu}.  On the other hand, the observation of a band gap $\Delta(U=0)\sim 0.6 $ eV in generalized gradient approximation (GGA) of DFT, led to the conclusion that the $X$-type {\material} is a magnetic semiconductor \cite{dft:K0.8Fe1.6Se2:Yan}. Hence, it is critical to clarify the $U$ dependence of $\Delta$.
%with Perdew-Burke-Ernzerhof formula for the exchange correlation potentials.

To this end, we performed a series of GGA+$U$ calculations for a number of $U$ on the Fe atoms, different spin orders, and Fe-Fe bond lengths, since a Mott gap generally scales up with $U$, independent of spin order, and is anticorrelated with the bond strength. As revealed in Fig.~3(a), the Fe partial density of states (DOS) shows that Fe $3d^6$ is always in the high-spin configuration (five spin-majority electrons and one spin-minority electron), manifesting the strong effect of Hund's rule coupling. It  exhibits two ``gaps'': One, a high-energy ``Mott gap'' between the spin-majority subbands and spin-minority subbands, which scales with $U$ and is insensitive to the spin order. Two, the low-energy real gap $\Delta$ resides at Fermi level (zero energy) within the spin-minority subbands.

The fact that $\Delta$ resides in one spin channel suggests that a direct way to decern its nature is to manipulate the bond length within the FM $2\times 2$ iron block (Fig.1f), and that a key clue is the experimentally observed tetramer lattice distortion (TLD) \cite{se122:xray:Zavalij}: The intra-block and inter-block Fe-Fe bond lengths are 2.691 and 2.916 {\AA}, respectively. We thus compared the $X$ type in the realistic structure and in a hypothetic TLD-free structure where all the Fe-Fe bond lengths are equal to 2.757 {\AA} and the Fe-Se bond lengths remain unchanged \cite{note:supplement}.

%\begin{figure*}[b]
\begin{figure}[t]
%\vspace{1cm}
\includegraphics[width=\columnwidth,clip=true,angle=0]{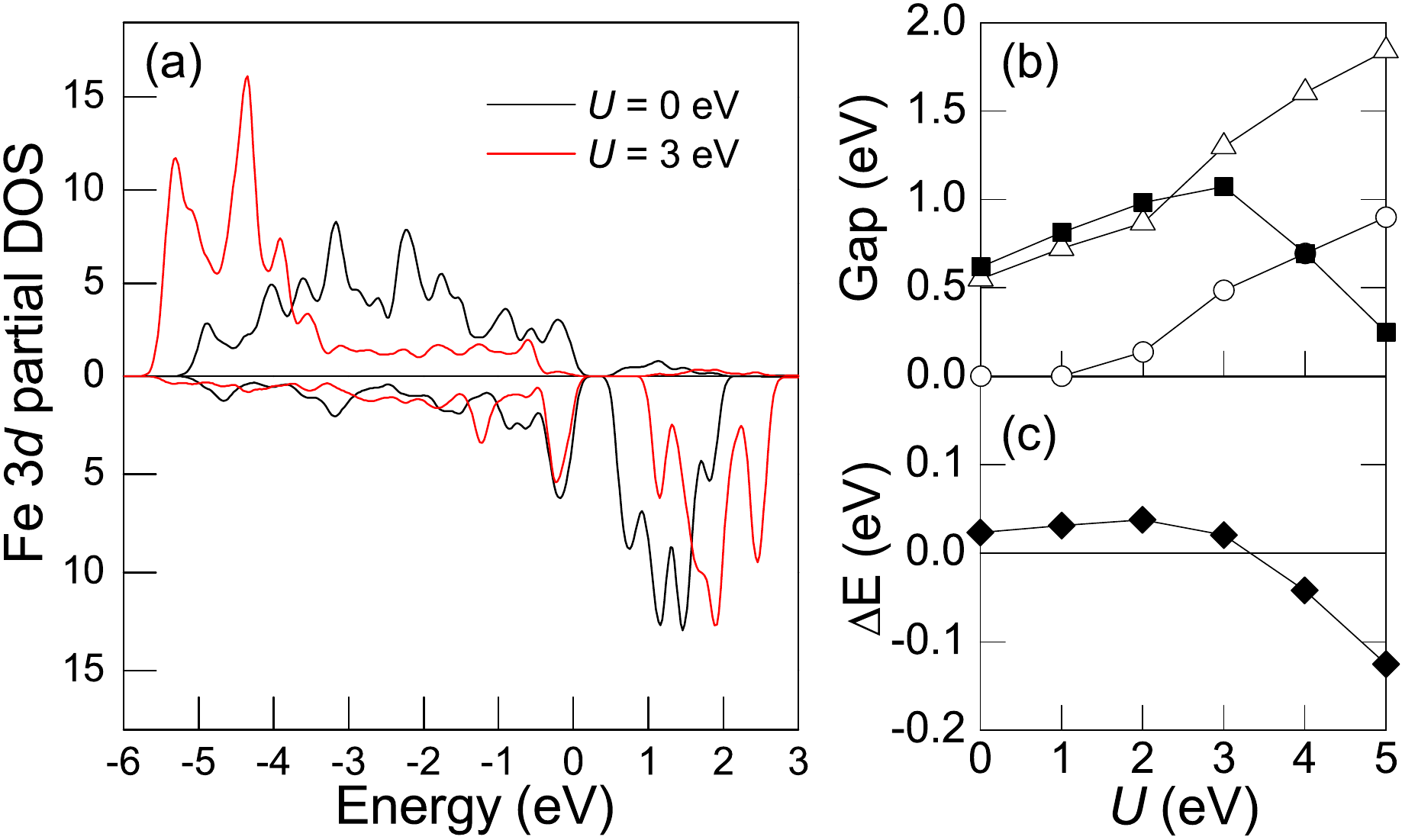}
\caption{\label{fig2:gap} (a) Fe $3d$ partial DOS of the \textit{X}-type AF state obtained from GGA+$U$ calculations with $U=0$ and $3$ eV. Upper and lower panels are for spin-majority and spin-minority, respectively. (b) The band gap size as a function of $U$ for $X$-type (solid squares) and FM (open circles) orders, as well as for the $X$-type without TLD (open triangles). (c) Total energy difference per Fe atom induced by vanishing TLD in the $X$ type as a function of $U$.}
%\caption{\label{fig2:gap} Band gap size of the \textit{X}-type AF ordered K$_{0.8}$Fe$_{1.6}$Se$_2$ (a) as a function of $U$ %in the first-principles LDA+$U$ method with two inputs for $J$, and (b) as a function of $KS$ in the orbital-degenerate DE %model.
%}
\end{figure}
%\end{figure*}

We found that the Mott insulator scenario disfavors the realistic structure. The total energy difference between these two structures as a function of $U$ is plotted in Fig.~3c. The realistic structure turns out to be unstable for $U>3$ eV. Consistently, as shown in Fig.~3b, the increase of $\Delta$ with $U$ is relatively slow in the realistic structure for $U<3$ eV (squares) but quick in the TLD-free structure for $U>3$ eV (triangles). Moreover, in the realistic regime ($U<3$ eV), $\Delta$ in the TLD-free structure is noticeably smaller than in the realistic case, indicating a positive correlation between $\Delta$ and intra-block Fe-Fe bond strength. Furthermore, $\Delta$ is especially sensitive to the magnetic structure for $U<3$ eV, as manifested by its vanishing values in the FM case (Fig.~3b). Hence, the realistic $\Delta$ results essentially from the bonding-antibonding splitting within the $2\times 2$ FM iron block.

\textsl{Effective Hamiltonian.---}The above first-principles results suggest the picture of low-energy itinerant electrons spin-polarized by Hund's rule coupling to more localized electronic states. A minimum model is the spin-fermion model where the Fe $d_{xz}$ and $d_{yz}$ orbitals were treated to host itinerant electrons and the rest Fe $3d$ orbitals were treated as an effective localized spin \cite{_DE:1111:Kou,_DE:Lv,_DE:11:Yin}:
\begin{eqnarray}
\label{eq1} H=&-&\sum\limits_{ij\gamma \gamma^\prime \mu }
{(t_{ij}^{\gamma \gamma^\prime} C_{i\gamma \mu }^\dag C_{j
\gamma^\prime \mu}^{} +h.c.)} \nonumber \\
&-& \frac{K}{2}\sum\limits_{i\gamma \mu \mu' } {C_{i\gamma \mu
}^\dag \vec {\sigma }_{\mu \mu' } C_{i\gamma \mu' }^{} }  \cdot \vec
{S}_i + \sum\limits_{ij} { J_{ij} \vec {S}_i \cdot \vec {S}_j},
\end{eqnarray}
where $C_{i\gamma \mu }^{} $ denotes the annihilation operator of an
itinerant electron with spin $\mu=\uparrow$ or $\downarrow $ in the
$\gamma=d_{xz}$ or $d_{yz} $ orbital on site $i$. $t_{ij}^{\gamma
\gamma^\prime} $'s are the electron hopping parameters. $\vec
{\sigma }_{\mu \mu' } $ is the Pauli matrix and $\vec {S}_i$ is the
localized spin whose magnitude is $S$. $K$ is the effective Hund's rule coupling. $J_{ij}$ is the AF
superexchange couplings between the localized spins; in particular,
$J$ and $J'$ are respectively the nearest-neighbor (NN) and
next-nearest-neighbor (NNN) ones. The filling of the itinerant electrons is on average three (one hole) per Fe site,  corresponding to the high-spin configuration of Fe $3d^6$ \cite{dft:1111:Lee}.

This model was proposed \cite{_DE:11:Yin} to unify the metallic \textit{C}-type and \textit{E}-type AF orders in vacancy-free FeSCs, assuming that $KS$ is the leading material-dependent parameter, controlled by the anion height from the iron plane; $KS$ was set to be $0.8$ eV for FeTe. To show the unifying capability of the model, below the \emph{same} set of parameters is used for {\material}. \ignore{That is, to the $y$ direction, the $d_{xz}$-$d_{xz}$ NN hopping integral $t_\| \simeq 0.4$ eV and the $d_{yz}$-$d_{yz}$ NN hopping integral $t_\bot \simeq 0.13$ eV; they are swapped to the $x$ direction; by symmetry the NN interorbital hoppings are zero; the NNN intraorbital hopping integral $t'\simeq -0.25$ eV for both $d_{xz}$ and $d_{yz}$ orbitals, and the NNN interorbital hopping is $\pm 0.07$ eV. $KS\simeq 0.8$ eV. $J S^2$ and $J' S^2$ are of the same order of $10$ meV. Since the Se anion is located above the center of the Fe plaquette, hybridizations via the Fe-As-Fe path are likely to give rise to comparable NN and NNN parameters. The filling of the itinerant electrons is on-average three per site, corresponding to the high-spin configuration of Fe $3d^6$.} The effect of TLD in {\material} is included via multiplying the intrablock (and interblock) hopping parameters by $1+\alpha$ (and $1-\alpha$), corresponding to the shrinking (and elongation) of the Fe-Fe bonds. Comparing the hopping parameters obtained from our first-principles Wannier function analysis \cite{dft:1111:Lee}, we found $\alpha \simeq 0.2$. The magnetic landscaping is studied by comparing a variety of static spin orders in the presence of ordered vacancies, such as the FM state and the AF states of \textit{C}-type (Fig.~1d), \textit{E}-type (Fig.~ 1e), \textit{X}-type (Fig.~1f), and \textit{G}-type (i.e., the N\'{e}el state where all NN spins are antiparallel), with the localized spins treated as Ising spins.

\begin{figure}[t]
%\vspace{1cm}
\includegraphics[width=\columnwidth,clip=true,angle=0]{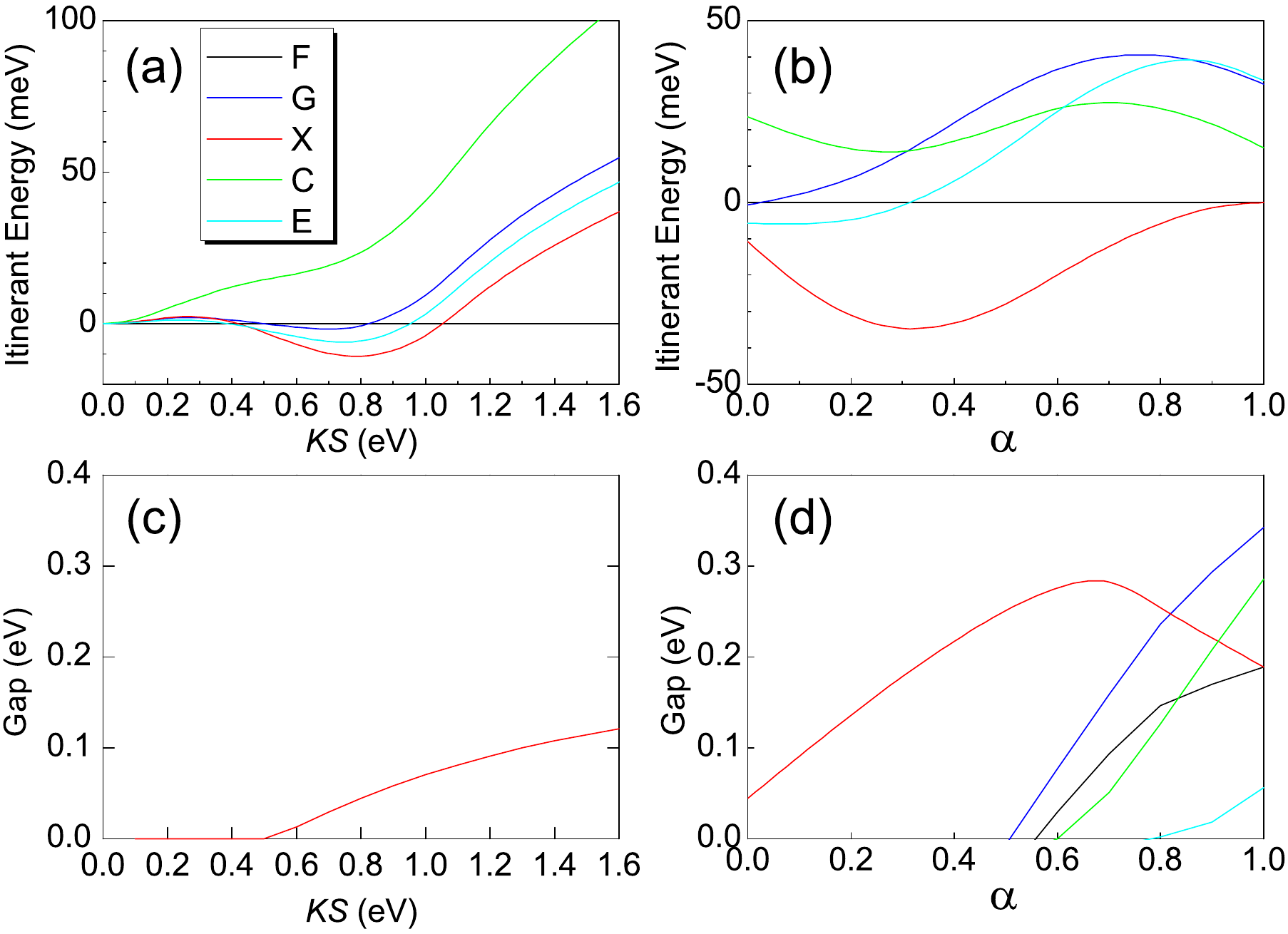}
\caption{\label{fig:phase}
(a) Itinerant energy per Fe versus $KS$ for the TLD parameter $\alpha=0$ (see text). The FM state is the energy reference. (b) Itinerant energy per Fe versus $\alpha$ for $KS=0.8$ eV. (c) Gap size versus $KS$ for $\alpha=0$. (d) Gap size versus $\alpha$ for $KS=0.8$ eV.
}
\end{figure}

Figs.~4a and 4b show how the $X$ type is stabilized. The localized-spin part of the model favors the \textit{C} type due to the comparable $J$ and $J'$ superexchange processes: Their contributions to the total energy per iron is $-1.5S^2J'$ for the \textit{C} type and $0.5S^2(J-J')$ for the \textit{X} type, for example. It is thus the itinerant energy, the energy involving the itinerant electrons, that favors the \textit{X} type around $KS=0.8$ eV (Fig.~4a). The itinerant energy is further lowered by TLD $\alpha$ remarkably  (Fig.~4b) enough to make the \textit{X} type more stable than the \textit{C} type for $J'S^2<30$ meV. These results agree well with the earlier \emph{ab initio} study of the TLD effect \cite{dft:K0.8Fe1.6Se2:Yan}. Furthermore, in the vicinity of the $X$ and $C$ types, the FM state is considerably higher in total energy than all these AF states by more than 50 meV per Fe atom (not shown), in agreement with the first-principles total energy calculations \cite{dft:K0.8Fe1.6Se2:Yan,dft:KFe1.6Se2:Cao}, a feature current spin-only model analysis failed to capture \cite{neu:se122:Wang,dft:KFe1.6Se2:Cao}. These results imply that with moderate $KS$, this system warrants strong and overall AF spin fluctuations in the Fe plane, providing a necessary environment for singlet superconductive electron pairing.

The insulating property of the $X$ type is also reproduced. The gap size as a function of $KS$ and $\alpha$ is shown in Figs.~4c and 4d, respectively. The nature of the band gap can be inferred from the limit of large $KS$, the so-called double-exchange limit where the electrons cannot hop to the sites with opposite spin orientation due to the energy barrier as high as $KS$. In this limit, the inter-block hoppings occur at distances longer than NNN (Fig.~1f). Thus, with the NN and NNN hopping parameters, any itinerant electron is localized within a $2\times 2$ block, leading to four discrete energy levels. Even for $KS=0.8$ eV, the inter-block hoppings are still strongly suppressed and cannot help develop overlapped bands from those energy levels, as predicted in Ref.~\cite{_DE:11:Yin} (FeTe is metallic because of unsuppressed NNN hoppings; see Fig.~1b). Therefore, the band gap originates from bonding-antibonding splitting in the spin-minority channel and its size is positively correlated with the intra-block Fe-Fe bonding strength represented by $\alpha$, indeed.

The above recognition that optimization of the Fe-Fe FM bonds is an important factor leads to the following intuitive insight into the vacancy-induced $E$-$X$ transition. As shown in Figs.~1b and 1f, both spin orders share a similar bond pattern: Each Fe atom is linked to one $d_{xz}$ FM bond and one $d_{yz}$ FM bond. We notice that another realization of this pattern is the vacancy-free $X$ type (a metallic $2\times 2$ block checkerboard AF order) shown in Fig.~1c. Thus, the vacancy-free $E$ and $X$ types are likely to have similar itinerant energy. Besides, their localized-spin part contributes exactly the same energy in the classic spin approximation. This implies that the vacancy-free $X$ and $E$ types could be very close in energy, in agreement with neutron scatter measurement on FeTe
\cite{neu:11:Zaliznyak}. Indeed, we confirmed that they differ by only 7 meV/Fe in \textit{ab initio} calculations and 3 meV/Fe in Eq. (1) for FeTe. Then, the added Fe vacancies break the bonding pattern in the \textit{E} type (Fig.~1e) but retain it in the rearranged \textit{X} type (Fig.~1f). This means that in iron chalcogenides, the $2\times 2$ block checkerboard AF order is already highly competitive in the absence of Fe vacancies, and emerges as the ground state upon introduction of Fe vacancies. This can also serve as the base to understand the insulating $2\times 2$ block AF order in BaFe$_2$Se$_3$ \cite{bafe2se3:krzton}.

It is noteworthy that for a system with degenerate orbitals, orbital ordering may occur. On-site orbital ordering was argued to drive the $C$ type in LaOFeAs \cite{dft:1111:Lee}, the $E$ type in FeTe \cite{_oo:Turner:11} and the $X$ type in {\material} \cite{_strong:K0.8Fe1.6Se2:Lv}. The aforementioned peculiar bond pattern implies that the on-site orbital polarization $P$ (the difference in the occupation numbers of the Fe $d_{xz}$ and $d_{yz}$ orbitals) is weak in {\material}. Our first-principles Wannier function analysis verified that $P \simeq 0$ in FeTe and $P \simeq 0.06$ in {\material}, much smaller than $P=0.17$ in LaOFeAs \cite{dft:1111:Lee}. The nonvanishing $P$ in {\material} is caused by the vacancy-imposed symmetry breaking, whose effect is surprisingly weak, confirming that the Fe-Fe bonding effect is dominant. Interestingly, while on-site orbital ordering is found weak (confirmed in Ref.~\cite{_weak:se122:Luo}), the bond pattern may be regarded as a bond orbital order.

The importance of the Hund's rule coupling $K$ in governing electronic correlation in the metallic state of vacancy-free FeSCs has also been recently demonstrated in LDA and dynamical mean-field theory, giving rise to a new term ``Hund's metal'' \cite{_strong:dmft:Yin,_strong:dmft:Yin_NM}. Likewise, insulating {\material} may be regarded as a realization of novel ``Hund's insulator'' in the sense that the band gap opens primarily to gain Hund's rule coupling energy rather than Hubbard repulsive energy (i.e., due to the blocking effect of $K$ rather than $U$). A measure to distinguish Hund's insulator and Mott insulator is checking whether the gap size is sensitive to the magnetic structure rather than $U$ or positively correlated with the FM bond strength. Thus, Hund's metal-insulator transition is expected to be much more sensitive to structural changes. This suggests that tuning the $A_{1-y}$Fe$_{2-x}$Se$_2$ materials through Hund's metal-insulator transition (e.g, by pressure) be an effective route to optimize their superconductive properties.

In summary, we have shown from first principles that the insulating $X$-type antiferromagnetism in vacancy-ordered {\material} is incompatible with a simple Fermi-surface nesting or Mott insulator scenario. We demonstrate that it can be unified with the metallic $C$-type and $E$-type antiferromagnetism of vacancy-free FeSCs in the spin-fermion model, where the competition between double-exchange ferromagnetism associated with Hund's rule coupling and superexchange antiferromagnetism is tuned by the ordered vacancies into a novel Hund's metal-insulator transition. These findings indicate that the blocking effects of Hund's rule coupling and the resulting electron correlation are crucial to the electronic and magnetic structures of FeSCs, and are likely at the heart of their high-$T_c$ mechanism.

%Acknowledgements
We thank G. Giovannetti and Z.-Y. Lu for helpful discussion and sharing their unpublished first-principles data. This work was supported by the U.S. Department of Energy (DOE), Office of Basic Energy Science, under Contract No. DE-AC02-98CH10886.

%\bibliography{Fe}
%merlin.mbs 2010-03-15 4.21a (PWD, AO, DPC)

\end{document}